\documentclass[a4paper]{article}

\usepackage[margin=1in]{geometry}

\usepackage[T1]{fontenc}
\newcommand{\changefont}[3]{
\fontfamily{#1} \fontseries{#2} \fontshape{#3} \selectfont}

\changefont{ptm}{m}{n}

\usepackage{setspace} 
 \usepackage{graphicx}    % needed for including graphics e.g. EPS, PS

\newcommand \be{\begin{equation}}
\newcommand \ee{\end{equation}}
\newcommand \ba{\begin{eqnarray}}
\newcommand \ea{\end{eqnarray}}

\def\bit{\begin{itemize}}
\def\eit{\end{itemize}}

\usepackage{amsfonts}
\usepackage{amssymb}
\usepackage{amsmath}
\usepackage{graphics}
\usepackage{mathrsfs}
\usepackage{color}

\newtheorem{theorem}{Theorem}[section]

\newtheorem{lemma}{Lemma}[section]

\long\def\symbolfootnote[#1]#2{\begingroup%
\def\thefootnote{\fnsymbol{footnote}}\footnote[#1]{#2}\endgroup} 

\begin{document}

%\begin{frontmatter}
%

\begin{center}
\Large \textbf{SICNNs with Li-Yorke Chaotic Outputs on a Time Scale}
\end{center}

\vspace{-0.3cm}
\begin{center}
\normalsize \textbf{Mehmet Onur Fen$^{a,} \symbolfootnote[1]{Corresponding Author Tel.: +90 312 365 9276,   E-mail: monur.fen@gmail.com}$, Fatma Tokmak Fen$^b$} \\
\vspace{0.2cm}
\textit{\textbf{\footnotesize$^a$Department of Mathematics, Middle East Technical University, 06800, Ankara, Turkey}} \\
\textit{\textbf{\footnotesize$^b$Department of Mathematics, Gazi University, 06500, Teknikokullar, Ankara, Turkey}}
\vspace{0.1cm}
\end{center}

\vspace{0.3cm}

\begin{center}
\textbf{Abstract}
\end{center}

\noindent\ignorespaces

In the present study, we investigate the existence of Li-Yorke chaos in the dynamics of shunting inhibitory cellular neural networks (SICNNs) on time scales. It is rigorously proved by taking advantage of external inputs that the outputs of SICNNs exhibit Li-Yorke chaos. The theoretical results are supported by simulations, and the controllability of chaos on the time scale is demonstrated by means of the Pyragas control technique. This is the first time in the literature that the existence as well as the control of chaos are provided for neural networks on time scales.

\vspace{0.2cm}
 
\noindent\ignorespaces \textbf{Keywords:} Shunting inhibitory cellular neural networks; Time scales; Li-Yorke chaos; Proximality; Frequent separation; Chaos control

\section{Introduction} \label{hnn_cyclic_intro}

Cellular neural networks (CNNs) have been paid much attention starting with the studies of Chua and Yang \cite{Chua1,Chua2}. Exceptional role in psychophysics, speech, perception, robotics, adaptive pattern recognition, vision, and image processing has been played by shunting inhibitory cellular neural networks (SICNNs), which was introduced by Bouzerdoum and Pinter \cite{bouzer1}. The reader is referred to \cite{bouzer2}-\cite{Cheung99} for the applications of SICNNs. Another subject that is also popular is the theory of time scales, which is first presented by Hilger \cite{Hilger88}. This theory has many applications in various scientific fields, and neural networks are no exception \cite{Bohner01}-\cite{Cheng15}.

The first mathematically rigorous definition of chaos was introduced by Li and Yorke \cite{Li75}. It was shown in \cite{Li75} that if a map on an interval has a point of period three, then it possesses chaos in the sense of Li-Yorke. The presence of an uncountable scrambled set is a distinguishing feature of the Li-Yorke chaos. It was indicated in \cite{Kuchta} that if a map on a compact interval has a two point scrambled set, then it possesses an uncountable scrambled set. Li-Yorke sensitivity, which links the Li-Yorke chaos with the notion of sensitivity, was studied by Akin and Kolyada \cite{Akin03}. It was proved in \cite{Blanchard02} that the presence of positive topological entropy implies chaos in the sense of Li-Yorke. According to Marotto \cite{Marotto78}, a multidimensional continuously differentiable map is Li-Yorke chaotic provided that it has a snap-back repeller. On the other hand, Marotto's Theorem was utilized by Li et al. \cite{PLi07} to demonstrate the existence of Li-Yorke chaos in a spatiotemporal chaotic system. Moreover, Li-Yorke chaos on several spaces in connection with the cardinality of its scrambled sets was studied within the scope of the paper \cite{Guirao05}, and generalizations of Li-Yorke chaos to mappings in Banach spaces and complete metric spaces were provided in \cite{Kloeden06}-\cite{Shi05}. The presence of Li-Yorke chaos in dynamic equations on time scales was first studied in the paper \cite{Akh4}. The extension mechanism of chaos in continuous-time systems can be found in the studies \cite{Akh5}-\cite{Akh_inpress}.

To the best of our knowledge, the existence of chaos has never been achieved for neural networks on time scales in the literature before. The main novelty of the present paper is the consideration of chaos for neural networks on time scales. We develop the concept of Li-Yorke chaos for SICNNs on time scales and prove its existence rigorously by making use of the reduction technique to impulsive differential equations, which was introduced by Akhmet and Turan \cite{Akhmet06}. The results are appropriate to obtain chaotic SICNNs on time scales with arbitrary high number of cells. Another novelty of the present study is the control of chaos on time scales. The Pyragas control method \cite{Pyragas92}-\cite{Zelinka} is utilized to control the chaos in the dynamics of SICNNs on time scales. Our results show that the Pyragas control method is suitable to control chaos not only in continuous-time or discrete-time systems, but also in systems on time scales.

It was revealed in the papers \cite{Freeman1,Freeman2} that chaotic dynamics is inevitable for brain activities, and it was suggested by Guevara et al. \cite{Guevara83} that chaotic behavior may be responsible for dynamical diseases such as schizophrenia, insomnia, epilepsy and dyskinesia. Besides, the presence of chaos in neural networks is useful for separating image segments \cite{Shibasaki12}, information processing \cite{Nara92,Nara95} and synchronization \cite{Shi14}-\cite{Jankowski94}, and it can improve the performance of CNNs on problems that have local minima in energy (cost) functions, since chaotic behavior of CNNs can help the network avoid local minima and reach the global optimum \cite{Suykens03}. Moreover, chaos is an important tool in CNNs for the studies of chaotic communication \cite{Cheng13}-\cite{Yifeng97} and combinatorial optimization problems \cite{Ohta03}.

Let us describe the model of SICNNs in its most original form \cite{bouzer1}. Consider a two dimensional grid of processing cells arranged into $m$ rows and $n$ columns, and let $C_{ij},$ $i=1,2,\ldots,m,$ $j=1,2,\ldots,n,$ denote the cell at the $(i,j)$ position of the lattice. In SICNNs, neighboring cells exert mutual inhibitory interactions of the shunting type. The dynamics of a cell $C_{ij}$ are described by the following nonlinear ordinary differential equation,
\begin{eqnarray} 
\begin{array}{l} \label{1}
x'_{ij}(t)=-a_{ij}x_{ij}(t)-\displaystyle \sum_{C_{hl}\in N_{r}(i,j)} C_{ij}^{hl}f(x_{hl}(t))x_{ij}(t) + L_{ij}(t),
\end{array}
\end{eqnarray}
where $x_{ij}$ is the activity of the cell $C_{ij};$ $L_{ij}(t)$ is the external input to the cell $C_{ij};$ the constant $a_{ij}>0$ represents the passive decay rate of the cell activity; $C_{ij}^{hl}\geq 0$ is the coupling strength of postsynaptic activity of the cell $C_{hl}$ transmitted to the cell $C_{ij};$ the activation function $f(x_{hl})$ is a positive continuous function representing the output or firing rate of the cell $C_{hl};$ and the $r-$neighborhood of the cell $C_{ij}$ is defined as
\begin{eqnarray*} 
N_{r}(i,j)=\{C_{hl}: \max(|h-i|,|l-j|)\leq r,  ~ \ 1\leq h \leq m, ~  1 \leq l \leq n \}.
\end{eqnarray*}

%Throughout the paper, we will denote by $\mathbb R,$ $\mathbb N$ and $\mathbb Z$ the sets of all real numbers, natural numbers and integers, respectively.  

Motivated by the deficiency of mathematical methods for the investigation of chaos in neural networks on time scales, in the present study, we will consider the following SICNN,
\begin{eqnarray} \label{SICNN_timescale}
x^{\Delta}_{ij}(t)= -a_{ij} x_{ij}(t) -\displaystyle \sum_{C_{hl}\in N_{r}(i,j)} C_{ij}^{hl} f(x_{hl}(t)) x_{ij}(t) + L_{ij}(t,\zeta), ~ t\in \mathbb T_0,
\end{eqnarray} 
where $\mathbb T_0 = \bigcup_{k=-\infty}^{\infty} [\theta_{2k-1}, \theta_{2k}]$ and the sequence $\left\{\theta_k\right\}_{k \in \mathbb Z}$ is strictly increasing such that $\left|\theta_k\right| \to \infty$ as $\left|k\right|\to \infty.$ The external input $L_{ij}(t,\zeta)$ is defined through the equation $L_{ij}(t,\zeta)=\zeta^{ij}_k$ for $t\in [\theta_{2k-1}, \theta_{2k}],$ $k\in\mathbb Z,$ such that the sequence $\zeta=\left\{\zeta_k\right\}_{k\in\mathbb Z},$ $\zeta_k= \left\{\zeta^{ij}_k\right\} \in \mathbb R^{mn},$ is  generated by the map 
\begin{eqnarray}\label{timescale_mapF}
\zeta_{k+1}=F(\zeta_k),
\end{eqnarray}
where $F:\Lambda \to \Lambda$ is a continuous function and $\Lambda$ is a compact subset of $\mathbb R^{mn}.$ We assume that $\sum_{-\infty} (\theta_{2k}-\theta_{2k-1}) =\infty$ and $\sum^{\infty} (\theta_{2k}-\theta_{2k-1})=\infty.$ In this paper, the chaotic dynamics of SICNN \eqref{SICNN_timescale} is investigated. We rigorously prove the existence of Li-Yorke chaos in the network \eqref{SICNN_timescale} provided that the map \eqref{timescale_mapF} is Li-Yorke chaotic. One of the significant features of the obtained chaos is its controllability. We also demonstrate the control of the obtained chaos by means of the Pyragas control technique \cite{Pyragas92}.

The dynamics of \eqref{SICNN_timescale} is essentially non-autonomous, and it is difficult to verify the ingredients of chaos for unspecified time scales. That is the reason why we utilize the time scale introduced by Akhmet and Turan \cite{Akhmet06,Akhmet09} as well as the reduction technique to impulsive differential equations \cite{Akhmet06}.

Artificial neural networks on time scales of the form $\mathbb T_0$ can be useful to model neural processes whose instantaneous rate of change cannot be expressed through a differential equation, i.e., not achievable for differential equation analysis. In other words, one can take into account the intervals $[\theta_{2k-1},\theta_{2k}],$ $k\in \mathbb Z,$ as the ones in which the neural processes are achievable so that the instantaneous rate (differential equation) can be obtained experimentally, and the intervals $[\theta_{2k},\theta_{2k+1}],$ $k\in \mathbb Z,$ can be considered as the ones in which the neural process cannot be achieved. On the other hand, neural processes on time scales which are unions of disjoint compact intervals can be interpreted as activities that work \textit{intermittently}, i.e., activities that proceed at certain disjoint time intervals and discontinue in between. This type of behavior can be observed in many biological neural activities. For instance, there exist hearth interneurons and motor neurons in the leech whose activities last for about 6 seconds and then the neurons become silent for about 6 seconds \cite{Dowling01}. Besides, in Xenopus tadpoles, fictive swimming activity can be initiated by brief touch and swimming normally stops when the tadpole contacts an object or the water surface with its head \cite{Roberts97}-\cite{Perrins02}. The role of inhibitory reticulospinal neurons for this behavior was investigated by Perrins et al. \cite{Perrins02}. In addition to these, intermittent external inputs are effective tools for the treatment of neurological disorders. It was demonstrated by Boggs et al. \cite{Boggs06} that intermittent electrical stimulation of the pudendal nerve can restore bladder emptying. Moreover, the potential of intermittent vagus nerve stimulation to treat chronic hypertension and cardiac arrhythmias was studied by Annoni et al. \cite{Annoni15}.

The existence and stability of periodic, almost periodic and anti-periodic solutions of neural networks on time scales have been extensively investigated in the literature \cite{Zhou11}-\cite{Cheng15}. In particular, the papers \cite{Li11b}-\cite{Li12} were concerned with the dynamics of SICNNs on time scales. The existence and global exponential stability of anti-periodic solutions of impulsive SICNNs with distributed delays on time scales were studied by Li and Shu \cite{Li11b} by means of the coincidence degree method and Lyapunov functionals. On the other hand, periodic solutions of SICNNs of neutral type with time-varying delays in the leakage term on time scales were considered in the paper \cite{Li14}. Moreover, the existence and asymptotic stability of almost periodic solutions for SICNNs with time-varying and continuously distributed delays on time scales are studied in \cite{Li12} without assuming the global Lipschitz conditions of activation functions. The concept of almost periodicity was also considered within the scope of the paper \cite{Akhmet06} by using the reduction technique to impulsive differential equations, and the results of \cite{Akhmet06} can be developed for neural networks on time scales. However, chaos in neural networks on time scales was not taken into account in these studies. In the papers \cite{Gui06,Sun07}, SICNNs with impulsive effects were investigated, and it was numerically demonstrated that chaos exists in such networks without a theoretical support. The presence of chaos in continuous-time models of SICNNs with impulses, delay and chaotic/almost periodic postsynaptic currents was rigorously proved and supported by simulations in \cite{Akhmet2013b}-\cite{Akh_inpress}. Contrarily, in the present paper, we rigorously prove the existence of chaos in SICNNs on time scales. The control of the chaos on time scales is another novelty of our study.

The rest of the paper is organized as follows. In Section \ref{sec_prelim_SICNN_time_scales}, some basic concepts about differential equations on time scales are provided, and sufficient conditions for the existence of chaos are given. We investigate the existence, uniqueness and attractiveness feature of the bounded solutions of the network \eqref{SICNN_timescale} in Section \ref{bounded_solns_sec}. The main result of the present study is indicated in Section \ref{chaotic_sec}, where we rigorously prove the presence of Li-Yorke chaos in the dynamics of \eqref{SICNN_timescale}. An illustrative example is presented in Section \ref{SICNN_timescale_example_sec}, and a chaos control technique is numerically demonstrated in Section \ref{control_SICNN_timescale} by means of the Pyragas control method. Finally, Section \ref{conc_section_SICNN_timescales} is devoted for conclusions.

\section{Preliminaries} \label{sec_prelim_SICNN_time_scales}

The basic concepts about differential equations on time scales are as follows \cite{Bohner01,Lakshmikantham96}. A time scale is a nonempty closed subset of $\mathbb{R}.$ On a time scale $\mathbb{T}$, the forward and backward jump operators are defined as $\sigma(t)=\inf\left\{s\in\mathbb{T}: s>t \right\}$ and $\rho(t)=\sup\left\{s\in\mathbb{T}: s<t\right\},$ respectively. A point $t\in\mathbb{T}$ is called right-scattered if $\sigma(t)>t$ and right-dense if $\sigma(t)=t$. Similarly, if $\rho(t)<t,$ then $t\in\mathbb{T}$ is called left-scattered, and otherwise it is called left-dense. We say that a function $\vartheta:\mathbb{T} \to \mathbb{R}^{mn},$ $\vartheta(t)=\left\{\vartheta_{ij}(t)\right\},$ $i=1,2,\ldots,m,$ $j=1,2,\ldots,n,$ is rd-continuous provided it is continuous at each right-dense point $t \in \mathbb{T},$ and its left-sided limits exist at each left-dense point $t\in \mathbb T.$ At a right-scattered point $t\in\mathbb{T},$ the $\Delta$-derivative of a continuous function $\vartheta$ is defined as $\displaystyle \vartheta^{\Delta}\left(t\right)=\frac{\vartheta\left(\sigma\left(t\right)\right)-\vartheta\left(t\right)}{\sigma\left(t\right)-t}.$ On the other hand, at a right-dense point $t,$ we have $\displaystyle \vartheta^{\Delta}\left(t\right)=\lim_{s \to t}\frac{\vartheta\left(t\right)-\vartheta\left(s\right)}{t-s}$ provided that the limit exists.

We suppose that the time scale $\mathbb T_0$ used in SICNN \eqref{SICNN_timescale} satisfies the $\omega-$property, i.e., there exists a positive number $\omega$ such that $t+\omega \in \mathbb T_0$ whenever $t \in \mathbb T_0.$ Under this assumption, there exists a natural number $p$ such that $\delta_{k+p}=\delta_k$ for all $k\in \mathbb Z,$ where $\delta_k=\theta_{2k+1}-\theta_{2k}$ \cite{Akhmet06}.  It is worth noting that the points $\theta_{2k-1}$, $k\in\mathbb{Z},$ are left-scattered and right-dense, and the points $\theta_{2k}$, $k\in\mathbb{Z},$ are right-scattered and left-dense on the time scale $\mathbb T_0.$ Moreover, $\sigma(\theta_{2k})=\theta_{2k+1}$, $\rho(\theta_{2k+1})=\theta_{2k}$, $k\in\mathbb{Z},$ and $\sigma(t)=\rho(t)=t$ for any $t\in\mathbb{T}_{0}$ except at the points $\theta_k,$ $k\in\mathbb Z.$

Let us assume without loss of generality that $\theta_{-1}<0<\theta_0,$ and define the $\psi-$substitution \cite{Akhmet06} on the set $\mathbb T'_0=\mathbb T_0 \setminus \bigcup_{k =-\infty}^{\infty} \left\{\theta_{2k-1} \right\}$ as
\begin{eqnarray} \label{func_psi}
\displaystyle \psi(t)=\left\{\begin{array}{ll} \displaystyle t-\sum_{0<\theta_{2k}<t} \delta_k,    &   t  \ge 0, \\
                                                 \displaystyle t+\sum_{t\le\theta_{2k}<0} \delta_k,    &   t < 0, 
\end{array} \right..
\end{eqnarray}
One can confirm that the function $\psi:\mathbb T'_0 \to \mathbb R$ is one-to-one and onto, $\psi(0)=0$ and $\displaystyle \lim_{t\to\infty, ~t\in \mathbb T_0'}\psi(t) = \infty.$ It was shown by Akhmet and Turan \cite{Akhmet06} that $d\psi(t)/dt=1,$ $t\in \mathbb T'_0,$ and $d\psi^{-1}(s)/ds=1$ provided that $s\neq s_k,$ $k\in\mathbb Z,$ where 
\begin{eqnarray} \label{func_psi_inv}
\displaystyle \psi^{-1}(s)=\left\{\begin{array}{ll} \displaystyle s+\sum_{0<s_{k}<s} \delta_k,    &   s  \ge 0, \\
                                                 \displaystyle s-\sum_{s\le s_{k}<0} \delta_k,    &   s < 0, 
\end{array} \right.
\end{eqnarray}
and the sequence $\left\{s_k\right\}_{k\in \mathbb Z}$ is defined through the equation $s_k=\psi(\theta_{2k})$ for each $k \in \mathbb Z.$ The function $\psi^{-1}$ is piecewise continuous with discontinuities of the first kind at the points $s_k,$ $k\in \mathbb Z,$ such that $\psi^{-1}(s_k+)-\psi^{-1}(s_k)=\delta_k,$ where $\displaystyle \psi^{-1}(s_k+)=\lim_{s\to s_k^+} \psi^{-1}(s).$ The sequence $\left\{s_k\right\}_{k\in \mathbb Z}$ is $(\psi(\omega),p)-$periodic, i.e., $s_{k+p}=s_k+\psi(\omega)$ for all $k\in\mathbb Z,$ and $\psi(t+\omega)=\psi(t)+\psi(\omega),$ $t\in \mathbb T'_0.$ Moreover, if a function $h(t)$ is $\omega-$periodic on $\mathbb T_0,$ then $h(\psi^{-1}(s))$ is $\psi(\omega)-$periodic, and vice versa.

One of the significant results mentioned in the paper \cite{Akhmet06} concerning the functions defined on $\mathbb R$ and $\mathbb T_0$ is as follows. Denote by ${\cal C}_{rd}(\mathbb T_0)$ the set of all rd-continuous functions $\vartheta:\mathbb T_0 \to \mathbb R^{mn},$ $\vartheta(t)=\left\{\vartheta_{ij}(t)\right\},$ and let ${\cal C}^1_{rd}(\mathbb T_0)\subset {\cal C}_{rd}(\mathbb T_0)$ be the set of all continuously differentiable functions on $\mathbb T_0,$ assuming that the functions have a one sided derivative at $\theta_k,$ $k\in\mathbb Z.$ Moreover, a function $\widetilde{\vartheta}:\mathbb R \to \mathbb R^{mn},$ $\widetilde{\vartheta}(t)=\left\{\widetilde{\vartheta}_{ij}(t)\right\},$ is an element of the set ${\cal PC}(\mathbb R)$ if it is left-continuous on $\mathbb R$ and continuous on $\mathbb R \setminus \bigcup_{k=-\infty}^{\infty}\left\{s_k\right\},$ and it has discontinuities of the first kind at the points $s_k,$ $k\in\mathbb Z.$ The function $\widetilde{\vartheta}$ belongs to the set ${\cal PC}^1(\mathbb R)$ if both $\widetilde{\vartheta}$ and $\widetilde{\vartheta}'$ are elements of ${\cal PC}(\mathbb R),$ where $\widetilde{\vartheta}'(s_k)= \displaystyle \lim_{s\to s_k^-} \frac{\widetilde{\vartheta}(s)-\widetilde{\vartheta}(s_k)}{s-s_k},$ $k\in \mathbb Z.$ It was proved in \cite{Akhmet06} that a function $\vartheta(t)$ belongs to ${\cal C}_{rd}(\mathbb T_0)$ $\left({\cal C}^1_{rd}(\mathbb T_0)\right)$ if and only if $\vartheta(\psi^{-1}(s))$ belongs to ${\cal PC}(\mathbb R)$ $\left({\cal PC}^1(\mathbb R)\right).$

Making use of the equation 
\begin{eqnarray*} \label{SICNN_paper_eqn1}
x_{ij}^{\Delta}(\theta_{2k}) = \displaystyle \frac{x_{ij}(\theta_{2k+1})-x_{ij}(\theta_{2k})}{\delta_k}, \ \  k \in \mathbb Z,
\end{eqnarray*} 
SICNN \eqref{SICNN_timescale} can be written as 
\begin{eqnarray} \label{transformed_system}
\begin{array}{l}
x'_{ij}(t) = - a_{ij} x_{ij}(t) - \displaystyle \sum_{C_{hl}\in N_{r}(i,j)} C_{ij}^{hl} f(x_{hl}(t)) x_{ij}(t) + L_{ij}(t,\zeta), ~ t\in \mathbb T_0, \\
x_{ij}(\theta_{2k+1}) =  (1  -\delta_ka_{ij})x_{ij}(\theta_{2k})  - \displaystyle  \sum_{C_{hl}\in N_{r}(i,j)} C_{ij}^{hl}  f(x_{hl}(\theta_{2k})) x_{ij}(\theta_{2k}) \delta_k   +\zeta_k^{ij} \delta_k.
\end{array}
\end{eqnarray}
Applying the transformation $s=\psi(t)$ to  (\ref{transformed_system}) we obtain the following impulsive network,
\begin{eqnarray} \label{impulsive_system}
\begin{array}{l}
y'_{ij}(s) = - a_{ij} y_{ij}(s) - \displaystyle \sum_{C_{hl}\in N_{r}(i,j)} C_{ij}^{hl} f(y_{hl}(s)) y_{ij}(s) + L_{ij}(\psi^{-1}(s),\zeta), ~ s\neq s_k, \\
\Delta y_{ij}|_{s=s_k} =  -\delta_k a_{ij} y_{ij}(s_{k})  - \displaystyle  \sum_{C_{hl}\in N_{r}(i,j)} C_{ij}^{hl}  f(y_{hl}(s_{k})) y_{ij}(s_{k}) \delta_k   +\zeta_k^{ij} \delta_k,
\end{array}
\end{eqnarray}
where $\Delta y_{ij}|_{s=s_k} = y_{ij}(s_k+) - y_{ij}(s_k)$ and $y_{ij}(s_k+)=\displaystyle \lim_{s \to s_k^+} y_{ij}(s),$ $k \in \mathbb Z.$

In the remaining parts of the paper, we will make use of the norm $\left\|\vartheta\right\|=\displaystyle \max_{(i,j)} \left|\vartheta_{ij}\right|,$ where $\vartheta=\left\{\vartheta_{ij}\right\} = (\vartheta_{11},\ldots,\vartheta_{1n}, \ldots, \vartheta_{m1} \ldots,\vartheta_{mn}) \in \mathbb R^{m n}.$  

The following conditions are required throughout the paper. 
\begin{enumerate}
\item[\textbf{(C1)}] $\delta_k a_{ij} \neq 1$ for all $i= 1,2,\ldots,m,$ $j= 1,2,\ldots,n$ and $k \in \mathbb Z;$
\item[\textbf{(C2)}] $\displaystyle \lambda= \min_{(i,j)} \lambda_{ij} > 0,$ where $\lambda_{ij} = a_{ij} - \displaystyle  \frac{1}{\psi(\omega)} \sum_{\nu=0}^{p-1} \ln \left|1-\delta_{\nu} a_{ij}\right|;$
\item[\textbf{(C3)}] There exists a positive number $M_f$ such that $\displaystyle \sup_{s \in \mathbb R}\left|f(s)\right|\le M_f;$ 
\item[\textbf{(C4)}] There exists a positive number $L_f$ such that $\left|f(s_1)-f(s_2)\right| \le L_f \left|s_1-s_2\right|$ for all $s_1,$ $s_2 \in \mathbb R.$
\end{enumerate}

Let us define $u_{ij}(s,\tau)= e^{-a_{ij} (s-\tau)} \prod_{\nu=l}^k (1-\delta_{\nu} a_{ij}),$ in the case that $s$ and $\tau$ are numbers such that $s_{l-1}< \tau \le s_l,$ $s_k < s \le s_{k+1}$ for some integers $k$ and $l$ with $k \ge l.$ On the other hand, if $s_{k}<\tau \le s\le s_{k+1}$ for some integer $k,$ then take $u_{ij}(s,\tau) = e^{-a_{ij}(s-\tau)}.$ One can verify under the conditions $(C1)$ and $(C2)$ that for each $i=1,2,\ldots,m,$ $j=1,2,\ldots,n,$ there exist positive numbers $K_{ij}$ such that $\left|u_{ij}(s,\tau)\right|\le K_{ij} e^{-\lambda_{ij}(s-\tau)},$ $s\ge \tau.$

Suppose that $M_f \overline{c} < 1,$ where $\overline{c}=\displaystyle \max_{(i,j)}\bigg(\frac{K_{ij}}{\lambda_{ij}}+\frac{p \delta K_{ij}}{1-e^{-\lambda_{ij} \psi(\omega)}}\bigg)\sum_{C_{hl}\in N_{r}(i,j)} C_{ij}^{hl}$ and $\displaystyle \delta=\max_{1\le k \le p} \delta_k.$ In the remaining parts of the paper, we will denote and $M_F= \displaystyle \max_{\eta \in \Lambda} \left\| F(\eta) \right\|,$ $H_0=\displaystyle \frac{M_F}{1-M_f \overline{c}}\max_{(i,j)} \bigg(\frac{K_{ij}}{\lambda_{ij}}+\frac{p \delta K_{ij}}{1-e^{-\lambda_{ij} \psi(\omega)}}\bigg)$ and $\displaystyle \overline{d}=(M_f+H_0L_f) \max_{(i,j)} K_{ij}\sum_{C_{hl}\in N_{r}(i,j)} C_{ij}^{hl}.$

The following conditions are also needed.
\begin{enumerate}
\item[\textbf{(C5)}] $(M_f+H_0L_f) \overline{c} < 1;$
\item[\textbf{(C6)}] $-\lambda + \overline{d} + \displaystyle \frac{p}{\psi(\omega)} \ln(1+\delta\overline{d} )<0.$
\end{enumerate}

In the next section, we will deal with the existence, uniqueness and attractiveness property of the bounded solutions of SICNN \eqref{SICNN_timescale}.

%%%%%%%%%%%%% new section %%%%%%%%%%%%%%%

\section{Bounded solutions} \label{bounded_solns_sec}

According to the results of \cite{Akh1,Samolienko95}, a bounded on $\mathbb R$ function $y(s)=\left\{y_{ij}(s)\right\},$ $i=1,2,\ldots,m,$ $j=1,2,\ldots,n,$ is a solution of the impulsive system (\ref{impulsive_system}) if and only if the equation
\begin{eqnarray*}
&& y_{ij}(s) = -\displaystyle \int_{-\infty}^{s} u_{ij}(s,\tau)\Big[\sum_{C_{hl}\in N_{r}(i,j)} C_{ij}^{hl} f(y_{hl}(\tau))y_{ij}(\tau)-L_{ij}(\psi^{-1}(\tau),\zeta)\Big] d\tau\\
&& -\displaystyle \sum_{-\infty<s_k<s} u_{ij}(s,s_k+)\Big[\sum_{C_{hl}\in N_{r}(i,j)} C_{ij}^{hl} f(y_{hl}(s_k))y_{ij}(s_k)-\zeta_k^{ij}\Big] \delta_k
\end{eqnarray*}
is valid.

The existence of bounded on $\mathbb R$ solutions of (\ref{impulsive_system}) is considered in the next assertion.

\begin{lemma} \label{boundedness_lemma_SICNN}
Suppose that the conditions $(C1)-(C5)$ are fulfilled. For each solution $\zeta=\left\{\zeta_{k}\right\}_{k \in \mathbb Z}$ of \eqref{timescale_mapF}, there exists a unique bounded on $\mathbb R$ solution $\phi_{\zeta}(s)=\left\{\phi_{\zeta}^{ij}(s)\right\},$ $i= 1,2,\ldots,m,$ $j= 1,2,\ldots,n$ of \eqref{impulsive_system} such that $\displaystyle \sup_{s \in \mathbb R}\left\|\phi_{\zeta}(s)\right\|\le H_0.$
\end{lemma}

\noindent \textbf{Proof.} Fix a solution $\zeta=\left\{\zeta_{k}\right\}_{k \in \mathbb Z}$ of \eqref{timescale_mapF} and take into account the set $\mathcal{D}_0 \subset PC(\mathbb R)$ consisting of functions of the form $z(s)=\left\{z_{ij}(s)\right\}$ which have discontinuities at the points $s_k,$ $k \in \mathbb Z$ such that $\left\|z\right\|_{\infty} \le H_0,$ where $\displaystyle \left\|z\right\|_{\infty}=\sup_{s \in \mathbb R} \left\|z(s)\right\|.$ The set $\mathcal{D}_0$ is complete \cite{Akh1}.

Define an operator $\Pi$ on $\mathcal{D}_0$  as 
\begin{eqnarray*}
&&(\Pi z(s))_{ij}= -\displaystyle \int_{-\infty}^{s} u_{ij}(s,\tau)\Big[\sum_{C_{hl}\in N_{r}(i,j)} C_{ij}^{hl} f(z_{hl}(\tau))z_{ij}(\tau)-L_{ij}(\psi^{-1}(\tau),\zeta)\Big] d\tau\\
&& -\displaystyle \sum_{-\infty<s_k<s} u_{ij}(s,s_k+)\Big[\sum_{C_{hl}\in N_{r}(i,j)} C_{ij}^{hl} f(z_{hl}(s_k))z_{ij}(s_k)-\zeta_k^{ij}\Big] \delta_k,
\end{eqnarray*}
where $\Pi z(s)=\left\{(\Pi z(s))_{ij}\right\}.$

If $z(s)\in \mathcal{D}_0,$ then one can confirm that
\begin{eqnarray*}
&& \left|(\Pi z(s))_{ij}\right| \le \displaystyle \int^s_{-\infty} K_{ij} e^{-\lambda_{ij}(s-\tau)} \Big( H_0 M_f \sum_{C_{hl}\in N_{r}(i,j)} C_{ij}^{hl} + M_F\Big) d\tau \\
&& + \displaystyle \sum_{-\infty < s_k < s} K_{ij} e^{-\lambda_{ij} (s-s_k)}  \Big( H_0 M_f \sum_{C_{hl}\in N_{r}(i,j)} C_{ij}^{hl} + M_F\Big) \delta \\
&& \le \bigg( \frac{K_{ij}}{\lambda_{ij}}  + \frac{p \delta K_{ij}}{1-e^{-\lambda_{ij} \psi(\omega)}} \bigg)  \Big( H_0 M_f \sum_{C_{hl}\in N_{r}(i,j)} C_{ij}^{hl} + M_F\Big).
\end{eqnarray*}
The last inequality implies that 
$\displaystyle \left\|\Pi z (s)\right\| \le H_0 M_f \overline{c} + M_F \max_{(i,j)}\bigg( \frac{K_{ij}}{\lambda_{ij}}  + \frac{p \delta K_{ij}}{1-e^{-\lambda_{ij} \psi(\omega)}} \bigg) = H_0.$
Therefore, $\Pi(\mathcal{D}_0) \subseteq \mathcal{D}_0.$

On the other hand, for any $z(s), \widetilde{z}(s) \in \mathcal{D}_0,$ we have
\begin{eqnarray} \label{SICNN_proof1}
\begin{array}{l}
 \left|(\Pi z(s))_{ij}-(\Pi \widetilde{z}(s))_{ij}\right| \leq  \displaystyle \int_{-\infty}^{s} M_f K_{ij} e^{-\lambda_{ij}(s-\tau)}  \sum_{C_{hl}\in N_{r}(i,j)} C_{ij}^{hl} \left| z_{ij}(\tau)-  \widetilde{z}_{ij}(\tau) \right| d\tau  \\
 + \displaystyle \int_{-\infty}^{s} H_0 L_f K_{ij} e^{-\lambda_{ij}(s-\tau)} \sum_{C_{hl}\in N_{r}(i,j)} C_{ij}^{hl}  \left| z_{hl}(\tau)-  \widetilde{z}_{hl}(\tau) \right| d\tau\\
 +\displaystyle \sum_{-\infty<s_k<s} M_f \delta K_{ij} e^{-\lambda_{ij} (s-s_k)} \sum_{C_{hl}\in N_{r}(i,j)} C_{ij}^{hl} \left|z_{ij}(s_k)-  \widetilde{z}_{ij}(s_k) \right|  \\
 +\displaystyle \sum_{-\infty<s_k<s} H_0 L_f \delta K_{ij} e^{-\lambda_{ij} (s-s_k)} \sum_{C_{hl}\in N_{r}(i,j)} C_{ij}^{hl} \left|z_{hl}(s_k)-  \widetilde{z}_{hl}(s_k) \right|. 
\end{array}
\end{eqnarray}
One can show by using \eqref{SICNN_proof1} that $\left\|\Pi z - \Pi \widetilde{z}\right\|_{\infty} \leq (M_f+H_0 L_f) \overline{c} \left\|z-\widetilde{z}\right\|_{\infty}.$ Thus, according to condition $(C5),$ the operator $\Pi$ is a contraction.
Consequently, there exists a unique bounded on $\mathbb R$ solution $\phi_{\zeta}(s)=\left\{\phi_{\zeta}^{ij}(s)\right\}$ of \eqref{impulsive_system} such that $\displaystyle \sup_{s \in \mathbb R}\left\|\phi_{\zeta}(s)\right\|\le H_0.$ $\square$

It is worth noting that for a given solution $\zeta=\left\{\zeta_{k}\right\}_{k \in \mathbb Z}$ of \eqref{timescale_mapF}, Lemma \ref{boundedness_lemma_SICNN} implies that the function $\varphi_{\zeta}(t)=\left\{\varphi_{\zeta}^{ij}(t)\right\}$ satisfying $\varphi_{\zeta}(t)=\phi_{\zeta}(\psi(t)),$ $t \in \mathbb T'_0,$ and $\varphi_{\zeta}(\theta_{2k+1})=\phi_{\zeta}(s_k+)$ is the unique solution of \eqref{SICNN_timescale}, which is bounded on $\mathbb T_0$ such that $\displaystyle \sup_{t \in \mathbb T_0}\left\|\varphi_{\zeta}(t)\right\|\le H_0.$

A bounded solution $\varphi_{\zeta}(t)$ of \eqref{SICNN_timescale} is said to attract another solution $x(t)=\left\{x_{ij}(t)\right\}$ of the same network if $\left\|x(t)-\varphi_{\zeta}(t)\right\| \to 0$ as $t \to \infty,$ $t \in \mathbb T_0.$ The next lemma is devoted to the attractiveness feature of the bounded solutions of \eqref{SICNN_timescale}. 

\begin{lemma}
If the conditions $(C1)-(C6)$ are valid, then for a fixed solution $\zeta=\left\{\zeta_{k}\right\}_{k \in \mathbb Z}$ of \eqref{timescale_mapF}, the bounded solution $\varphi_{\zeta}(t)$ of \eqref{SICNN_timescale} attracts all other solutions of the network.
\end{lemma}

\noindent \textbf{Proof.} Consider an arbitrary solution $x(t)=\left\{x_{ij}(t)\right\}$ of \eqref{SICNN_timescale} such that $x(t^0)=x_0,$ where $t^0 \in \mathbb T_0$ and $x_0=\left\{x^{ij}_0\right\} \in \mathbb R^{mn}.$ Assume without loss of generality that $t^0 \neq \theta_{2k-1}$ for any $k\in \mathbb Z,$ and let $y(s)=x(\psi^{-1}(s))$ and $s^0=\psi(t^0).$ In this case, for $s\ge s^0,$ the solutions $y(s)=\left\{y_{ij}(s)\right\}$ and $\phi_{\zeta}(s)=\left\{\phi_{\zeta}^{ij}(s)\right\}$ of \eqref{impulsive_system} satisfy the relation
\begin{eqnarray*}
&& y_{ij}(s) - \phi_{\zeta}^{ij}(s) = u_{ij}(s,s^0) \big( x_0^{ij} - \phi_{\zeta}^{ij}(s^0) \big)\\
&& - \displaystyle \int^s_{s^0} u_{ij}(s,\tau) \sum_{C_{hl}\in N_{r}(i,j)} C_{ij}^{hl}  \big[ f(y_{hl}(\tau))y_{ij}(\tau) - f(\phi_{\zeta}^{hl}(\tau)) \phi_{\zeta}^{ij}(\tau)  \big] d \tau \\
&& - \displaystyle \sum_{s^0 \le s_k < s} u_{ij}(s,s_k+) \sum_{C_{hl}\in N_{r}(i,j)} C_{ij}^{hl}  \big[ f(y_{hl}(s_k))y_{ij}(s_k) - f(\phi_{\zeta}^{hl}(s_k)) \phi_{\zeta}^{ij}(s_k)  \big] \delta_k.
\end{eqnarray*}
It can be verified by using the last equation that 
\begin{eqnarray*}
&& \left\|y(s)-\phi_{\zeta}(s)\right\| \le K e^{-\lambda (s-s^0)} \left\|x_0-\phi_{\zeta}(s^0)\right\| + \displaystyle \int^s_{s^0} \overline{d} e^{-\lambda (s-\tau)} \left\|y(\tau)-\phi_{\zeta}(\tau)\right\| d\tau \\
&& + \displaystyle \sum_{s^0 \le s_k < s} \delta \overline{d} e^{-\lambda(s-s_k)} \left\|y(s_k)-\phi_{\zeta}(s_k)\right\|, 
\end{eqnarray*}
where $\displaystyle K=\max_{(i,j)} K_{ij}.$
Benefiting from the Gronwall-Bellman Lemma for piecewise continuous functions \cite{Akh1}, we obtain for $s\ge s^0$ that
\begin{eqnarray*}
\left\|y(s)-\phi_{\zeta}(s)\right\| \le K (1+\delta \overline{d})^p \left\|x_0-\phi_{\zeta}(s^0)\right\| e^{[-\lambda + \overline{d} + p\ln(1+\delta \overline{d}) / \psi(\omega)]}(s-s^0).
\end{eqnarray*}
Condition $(C6)$ implies that $\left\|y(s)-\phi_{\zeta}(s)\right\| \to 0$ as $s \to \infty.$ Hence, $\left\|x(t)-\varphi_{\zeta}(t)\right\| \to 0$ as $t \to \infty,$ $t \in \mathbb T_0.$ $\square$

The presence of Li-Yorke chaos in SICNN \eqref{SICNN_timescale} will be investigated in the next section.

\section{Li-Yorke chaos} \label{chaotic_sec}

The map \eqref{timescale_mapF} is called Li-Yorke chaotic on $\Lambda$ if \cite{Li75,Akin03}: 
(i) For every natural number $p_0,$ there exists a $p_0-$periodic point of $F$ in $\Lambda;$ 
(ii) There is an uncountable set ${\mathcal S} \subset \Lambda,$ the scrambled set, containing no periodic points, such that for every $\xi_1=\left\{\xi_1^{ij}\right\},$ $\xi_2 =\left\{\xi_2^{ij}\right\}\in {\mathcal S}$ with $\xi_1 \neq \xi_2,$ we have $\displaystyle \limsup_{k\to\infty} \left\| F^k(\xi_1)-F^k \left(\xi_2 \right) \right\|>0$ and $\displaystyle \liminf_{k\to\infty} \left\|F^k(\xi_1)-F^k(\xi_2) \right\|=0;$ 
(iii) For every $\xi_1\in {\mathcal S}$ and a periodic point $\xi_2 \in \Lambda,$ we have $\displaystyle \limsup_{k\to\infty} \left\|F^k(\xi_1)-F^k(\xi_2)\right\| >0.$

Denote by $\Theta$ the set of all sequences $\zeta=\left\{\zeta_k\right\}_{k\in \mathbb Z}$ obtained by equation \eqref{timescale_mapF}. We say that a pair of sequences $\zeta=\left\{\zeta_k\right\}_{k \in \mathbb Z},$ $\tilde{\zeta}= \big\{\tilde{\zeta}_k\big\}_{k \in \mathbb Z} \in\Theta$ is proximal if $\displaystyle \liminf_{k\rightarrow\infty} \big\|\zeta_{k}-\tilde{\zeta}_{k}\big\| =0,$ and it is frequently separated if $\displaystyle \limsup_{k\rightarrow\infty} \big\|\zeta_{k}-\tilde{\zeta}_{k} \big\|>0.$

Let $\mathscr{A}$ be the collection of all bounded solutions $\varphi_{\zeta}(t)=\left\{\varphi^{ij}_{\zeta}(t)\right\}$ of \eqref{SICNN_timescale} such that $\zeta \in \Theta.$  A pair $\varphi_{\zeta}(t),$ $\varphi_{\widetilde{\zeta}}(t) \in \mathscr{A}$ is proximal if for an arbitrary small positive number $\epsilon$ and an  arbitrary large natural number $E,$ there exists an integer $k_0$ such that  $\big\|\varphi_{\zeta}(t) - \varphi_{\widetilde{\zeta}}(t)\big\|< \epsilon$ for all $t \in [\theta_{2k_{0}-1}, \theta_{2(k_0+E)}] \cap \mathbb T_0.$
On the other hand, the pair $\varphi_{\zeta}(t),$ $\varphi_{\widetilde{\zeta}}(t) \in \mathscr{A}$ is frequently $(\epsilon_0,\epsilon_1)$-separated if there exist numbers $\epsilon_0>0,$ $\epsilon_1>0$ and infinitely many disjoint intervals $J_q \subset \mathbb T_0,$ $q\in\mathbb N,$ each with a length no less than $\epsilon_1$ such that $\big\|\varphi_{\zeta}(t)-\varphi_{\widetilde{\zeta}}(t)\big\|>\epsilon_0$ for each $t$ from these intervals. Moreover, $\varphi_{\zeta}(t)$, $\varphi_{\widetilde{\zeta}}(t) \in \mathscr{A}$ is a Li-Yorke pair if it is proximal and frequently $(\epsilon_{0},\epsilon_1)$-separated for some positive numbers $\epsilon_{0}$ and $\epsilon_1.$

The SICNN \eqref{SICNN_timescale} is called Li-Yorke chaotic if:
(i) There exists a $p_0 \omega-$periodic solution of \eqref{SICNN_timescale} for each $p_0 \in \mathbb N;$  
(ii) There exists an uncountable set $\Sigma\subset \mathscr{A},$ the scrambled set, which does not contain any periodic solution,  such that any pair of different solutions of \eqref{SICNN_timescale} inside $\Sigma$ is a Li-Yorke pair; 
(iii) For any $\varphi_{\zeta}(t)\in \Sigma$ and any periodic solution $\varphi_{\widetilde{\zeta}}(t)\in \mathscr{A},$ the pair $\varphi_{\zeta}(t)$, $\varphi_{\widetilde{\zeta}}(t)$ is frequently $(\epsilon_{0},\epsilon_1)$-separated for some positive numbers $\epsilon_{0}$ and $\epsilon_1.$

One can confirm that the sequence $\left\{\eta_k\right\}$ is $p-$periodic, where $\eta_k=\theta_{2k}-\theta_{2k-1},$ $k\in \mathbb Z.$ Suppose that $\displaystyle \underline{\eta}=\min_{1\le k \le p} \eta_k$ and  $\displaystyle \overline{\eta}=\max_{1\le k \le p} \eta_k.$  We will denote by $i(J)$ the number of the terms of the sequence $\left\{s_k\right\}_{k \in \mathbb Z}$ that belong to the interval $J \subset \mathbb R.$ It can be verified that $i([a_0,b_0)) \le p + \displaystyle \frac{p}{\psi(\omega)} (b_0-a_0)$ for any real numbers $a_0,$ $b_0$ such that $b_0 > a_0.$

The proximality feature of bounded solutions of SICNN \eqref{SICNN_timescale} will be mentioned in the following lemma.

\begin{lemma} \label{timescale_SICNN_prox}
Assume that the conditions $(C1)-(C6)$ are satisfied. If a pair of sequences $\zeta,\tilde{\zeta}\in\Theta$ is proximal, then the same is true for the pair $\varphi_{\zeta}(t), \varphi_{\tilde{\zeta}}(t)\in\mathscr{A}.$
\end{lemma}

\noindent  \textbf{Proof.}
Throughout the proof, let us denote $$\alpha= \lambda - \overline{d} - \displaystyle \frac{p}{\psi(\omega)} \ln(1+\delta\overline{d}),$$
$$\beta_1 = 2 \max_{(i,j)} \Big( H_0 M_f \sum_{C_{hl}\in N_{r}(i,j)} C_{ij}^{hl} +M_F \Big) \bigg(\frac{K_{ij}}{\lambda_{ij}}+\frac{p \delta K_{ij}}{1-e^{-\lambda_{ij} \psi(\omega)}}\bigg),$$
and 
$$\beta_2 = \max_{(i,j)} \bigg(\frac{K_{ij} }{\lambda_{ij}}+\frac{p \delta K_{ij} }{1-e^{-\lambda_{ij} \psi(\omega)}}\bigg).$$
Assume that $\gamma$ is a real number which satisfies the inequality 
$$
\gamma \ge \displaystyle 1+ \beta_2 \left(1+ \frac{\overline{d}(1+\delta \overline{d})^p}{\alpha} + \frac{p \delta \overline{d}(1+ \delta \overline{d})^p }{1-e^{-\alpha \psi(\omega)}} \right).
$$
Fix an arbitrary small number $\epsilon>0$ and an arbitrary large natural number $E$ such that 
$$
\displaystyle E\ge \frac{1}{\alpha \underline{\eta}} \ln\left( \frac{\gamma \beta_1(1+ \delta \overline{d})^p}{\epsilon} \right).
$$ 
Since the pair $\zeta,$ $\tilde{\zeta}$ is proximal, there exists an integer $k_1$ such that   
$$
\left\| L \big(\psi^{-1}(s),\zeta \big) - L \big(\psi^{-1}(s),\widetilde{\zeta} \big) \right\| < \displaystyle\frac{\epsilon}{\gamma}
$$
for $s\in (s_{k_1-1}, s_{k_1+2E}].$

The bounded solutions $\phi_{\zeta}(s)=\left\{\phi^{ij}_{\zeta}(s)\right\},$ $\phi_{\widetilde{\zeta}}(s)=\left\{\phi^{ij}_{\widetilde{\zeta}}(s)\right\}$ of \eqref{impulsive_system} satisfy the relation
\begin{eqnarray*}
&& \phi^{ij}_{\zeta}(s) - \phi^{ij}_{\widetilde{\zeta}}(s) = - \displaystyle \int_{-\infty}^s u_{ij} (s,\tau) \Big[ \sum_{C_{hl}\in N_{r}(i,j)} C_{ij}^{hl} f \big( \phi_{\zeta}^{hl} (\tau) \big) \phi_{\zeta}^{ij}(\tau)- L_{ij} \big( \psi^{-1} (\tau), \zeta \big) \\
&& - \sum_{C_{hl}\in N_{r}(i,j)} C_{ij}^{hl} f \big( \phi_{\widetilde{\zeta}}^{hl} (\tau) \big) \phi_{\widetilde{\zeta}}^{ij}(\tau) + L_{ij} \big( \psi^{-1} (\tau), \widetilde{\zeta} \big)   \Big] d \tau \\ 
&& - \displaystyle \sum_{-\infty < s_k < s} u_{ij} (s, s_k+) \Big[ \sum_{C_{hl}\in N_{r}(i,j)} C_{ij}^{hl} f \big( \phi_{\zeta}^{hl} (s_k) \big) \phi_{\zeta}^{ij}(s_k)- \zeta_k^{ij} \\
&& - \sum_{C_{hl}\in N_{r}(i,j)} C_{ij}^{hl} f \big( \phi_{\widetilde{\zeta}}^{hl} (s_k) \big) \phi_{\widetilde{\zeta}}^{ij}(s_k)  + \widetilde{\zeta}_k^{ij}    \Big] \delta_k.   
\end{eqnarray*}
Therefore, for $s \in (s_{k_1-1}, s_{k_1+2E}],$ we have that
\begin{eqnarray} \label{time_scale_SICNN_proof1}
\begin{array}{l}
 \displaystyle  \left| \phi^{ij}_{\zeta}(s) - \phi^{ij}_{\widetilde{\zeta}}(s) \right| \leq 2 \Big( H_0 M_f \sum_{C_{hl}\in N_{r}(i,j)} C_{ij}^{hl} +M_F \Big) \displaystyle \bigg( \frac{K_{ij}}{\lambda_{ij}}+\frac{p \delta K_{ij}}{1-e^{-\lambda_{ij} \psi(\omega)}} \bigg) e^{- \lambda_{ij} (s-s_{k_1-1})} \\  
  + \displaystyle \frac{\epsilon K_{ij}}{\gamma \lambda_{ij}} \left( 1- e^{-\lambda_{ij} (s-s_{k_1-1})} \right) + \displaystyle \frac{p \delta \epsilon K_{ij}}{\gamma (1-e^{-\lambda_{ij} \psi(\omega)})} \left( 1- e^{-\lambda_{ij} (s-s_{k_1-1} + \psi(\omega))} \right) \\  
  + \displaystyle \int^s_{s_{k_1-1}} K_{ij} \sum_{C_{hl}\in N_{r}(i,j)} C_{ij}^{hl} (M_f + H_0 L_f) e^{-\lambda_{ij} (s-\tau)}  \left\| \phi_{\zeta} (\tau) - \phi_{\widetilde{\zeta}} (\tau)  \right\| d\tau \\  
  + \displaystyle \sum_{s_{k_1-1} < s_k < s}  K_{ij} \sum_{C_{hl}\in N_{r}(i,j)} C_{ij}^{hl} \delta (M_f + H_0 L_f) e^{-\lambda_{ij} (s-s_k)} \left\| \phi_{\zeta} (s_k) - \phi_{\widetilde{\zeta}} (s_k)  \right\|.
\end{array}
\end{eqnarray}

Define the function $v(s) = e^{\lambda s} \displaystyle \left\| \phi_{\zeta}(s) - \phi_{\widetilde{\zeta}}(s)\right\|.$ The inequality \eqref{time_scale_SICNN_proof1} implies that
\begin{eqnarray*}
v(s) \le \beta_1 e^{\lambda s_{k_1-1}} + \displaystyle \frac{\beta_2 \epsilon}{\gamma} e^{\lambda s} + \displaystyle \int_{s_{k_1-1}}^s \overline{d} v(\tau) d \tau + \sum_{s_{k_1-1} < s_k < s} \delta \overline{d} v(s_k), \ s \in (s_{k_1-1}, s_{k_1-1 + 2E}].
\end{eqnarray*}
Making use of the Gronwall's Lemma for piecewise continuous functions, one can obtain that
\begin{eqnarray*}
&& v(s) \le \beta_1 e^{\lambda s_{k_1-1}} (1+ \delta \overline{d})^{i((s_{k_1-1},s))}  e^{\overline{d} (s-s_{k_1-1})} + \displaystyle \frac{\beta_2 \epsilon}{\gamma} e^{\lambda s} \\
&& + \displaystyle \frac{\beta_2 \overline{d} \epsilon}{\gamma} \int^s_{s_{k_1-1}} (1+\delta \overline{d})^{i((\tau,s))} e^{\overline{d} (s-\tau)} e^{\lambda \tau} d\tau \\
&& + \displaystyle \frac{\beta_2 \delta \overline{d} \epsilon}{\gamma} \sum_{s_{k_1-1} < s_k < s}  (1+\delta \overline{d})^{i((s_k,s))} e^{\overline{d} (s-s_k)} e^{\lambda s_k}.
\end{eqnarray*}
Since the inequality 
$
(1+\delta \overline{d})^{i((a_0,b_0))} e^{\overline{d} (b_0-a_0)} \le (1+ \delta \overline{d})^p e^{(\lambda -\alpha)(b_0-a_0)} 
$
is valid for any real numbers $a_0$ and $b_0$ with $b_0 > a_0,$ we have 
\begin{eqnarray*}
&& v(s) \le \beta_1 e^{\lambda s_{k_1-1}} (1+\delta \overline{d})^p e^{(\lambda - \alpha)(s-s_{k_1-1})} + \displaystyle \frac{\beta_2\epsilon}{\gamma} e^{\lambda s} \\
&& + \displaystyle \frac{\beta_2 \overline{d} (1+\delta \overline{d})^p \epsilon}{\alpha \gamma} e^{\lambda s} \left( 1- e^{-\alpha (s-s_{k_1-1})} \right) 
 + \displaystyle \frac{\beta_2 p \delta \overline{d} (1+\delta \overline{d})^p \epsilon}{\gamma (1-e^{-\alpha \psi(\omega)})} e^{\lambda s}  \left( 1- e^{-\alpha (s-s_{k_1-1}+\psi(\omega))} \right).
\end{eqnarray*}
Hence, it can be verified that
$$
\displaystyle \left\|\phi_{\zeta}(s) - \phi_{\widetilde{\zeta}}(s) \right\| < \beta_1 (1+\delta \overline{d})^p e^{-\alpha(s-s_{k_1-1})}
+ \displaystyle \frac{\beta_2 \epsilon}{\gamma} \bigg(1+ \frac{\overline{d} (1+\delta \overline{d})^p}{\alpha} + \frac{p \delta \overline{d} (1+\delta \overline{d})^p}{1-e^{-\alpha \psi(\omega)}} \bigg).
$$
For $s\in(s_{k_1-1+E},s_{k_1+2E}],$ utilizing the inequality
$
\beta_1 (1+\delta \overline{d})^p e^{-\alpha (s-s_{k_1-1})} < \beta_1 (1+\delta \overline{d})^p e^{-\alpha E \underline{\eta}} \le \displaystyle \frac{\epsilon}{\gamma}
$
we attain that
$$
\left\|\phi_{\zeta}(s) - \phi_{\widetilde{\zeta}}(s)\right\| < \displaystyle \frac{\epsilon}{\gamma} \left[\displaystyle 1+ \beta_2 \left(1+ \frac{\overline{d}(1+\delta \overline{d})^p}{\alpha} + \frac{p \delta \overline{d}(1+ \delta \overline{d})^p }{1-e^{-\alpha \psi(\omega)}} \right)\right] \le \epsilon.
$$

Thus, $\left\|\varphi_{\zeta}(t)- \varphi_{\widetilde{\zeta}}(t)\right\|<\epsilon$ for $t \in [\theta_{2(k_1+E)-1}, \theta_{2(k_1+2E)}] \cap \mathbb T_0.$
Consequently, the pair $\varphi_{\zeta}(t), \varphi_{\tilde{\zeta}}(t)$ is proximal. $\square$

%%%%%%%%%%%%%%%% end of the proof %%%%%%%%%%%%%

The next lemma is devoted to the frequent separation feature in SICNN \eqref{SICNN_timescale}.

\begin{lemma} \label{SICNN_freq_separation_lemma}
Under the conditions $(C1)-(C5),$ if a pair of sequences $\zeta, \widetilde{\zeta} \in \Theta$ is frequently separated, then the pair of bounded solutions $\varphi_{\zeta}(t),$ $\varphi_{\widetilde{\zeta}}(t) \in \mathscr{A}$ is frequently $(\epsilon_0,\epsilon_1)$-separated for some positive numbers $\epsilon_0$ and $\epsilon_1.$
\end{lemma}

\noindent \textbf{Proof.} Since the pair $\zeta,$ $\widetilde{\zeta}$ is frequently separated, there exists a positive number $\overline{\epsilon}_0$ and a sequence $\left\{k_q\right\}$ of integers satisfying $k_q \to \infty$ as $q \to \infty$ such that $\left\|\zeta_{k_q} - \widetilde{\zeta}_q\right\| > \overline{\epsilon}_0,$ $q \in \mathbb N.$

Fix a natural number $q.$ If $s\in (s_{k_q-1},s_{k_q}],$ then the bounded solutions $\phi_{\zeta}(s)=\left\{\phi^{ij}_{\zeta}(s)\right\}$ and $\phi_{\widetilde{\zeta}}(s)=\left\{\phi^{ij}_{\widetilde{\zeta}}(s)\right\}$ of \eqref{impulsive_system} satisfy the relation
\begin{eqnarray*}
&& \phi^{ij}_{\zeta}(s)-\phi^{ij}_{\widetilde{\zeta}}(s) = \phi^{ij}_{\zeta}(s_{k_q-1}+)-\phi^{ij}_{\widetilde{\zeta}}(s_{k_q-1}+) - \displaystyle \int_{s_{k_q-1}}^{s} a_{ij} \left(\phi^{ij}_{\zeta}(\tau)-\phi^{ij}_{\widetilde{\zeta}}(\tau)\right) d\tau \\
&&  - \displaystyle \int_{s_{k_q-1}}^{s}  \sum_{C_{hl}\in N_{r}(i,j)} C_{ij}^{hl} \left[ f\big( \phi_{\zeta}^{hl} (\tau) \big) \phi_{\zeta}^{ij}(\tau) - f\big( \phi_{\widetilde{\zeta}}^{hl} (\tau) \big) \phi_{\widetilde{\zeta}}^{ij}(\tau)  \right] d\tau + \displaystyle \int_{s_{k_q-1}}^{s} \left( \zeta_{k_q}^{ij} - \widetilde{\zeta}_{k_q}^{ij}  \right) d\tau.
\end{eqnarray*}
Thus, it can be verified that
\begin{eqnarray*}
&& \big|\phi^{ij}_{\zeta}(s_{k_q})-\phi^{ij}_{\widetilde{\zeta}}(s_{k_q})\big| \ge \big| \zeta_{k_q}^{ij} - \widetilde{\zeta}_{k_q}^{ij} \big|   \left(  s_{k_q} - s_{k_q-1}\right) \\
&& - \big|\phi^{ij}_{\zeta}(s_{k_q-1}+)-\phi^{ij}_{\widetilde{\zeta}}(s_{k_q-1}+)\big| - \displaystyle \int_{s_{k_q-1}}^{s_{k_q}} a_{ij} \big|\phi^{ij}_{\zeta}(\tau)-\phi^{ij}_{\widetilde{\zeta}}(\tau)\big| d\tau \\
&&  - \displaystyle \int_{s_{k_q-1}}^{s_{k_q}}  \sum_{C_{hl}\in N_{r}(i,j)} C_{ij}^{hl} \big| f\big( \phi_{\zeta}^{hl} (\tau) \big) \phi_{\zeta}^{ij}(\tau) - f\big( \phi_{\widetilde{\zeta}}^{hl} (\tau) \big) \phi_{\widetilde{\zeta}}^{ij}(\tau)  \big| d\tau \\
&& \ge \big| \zeta_{k_q}^{ij} - \widetilde{\zeta}_{k_q}^{ij} \big|  \underline{\eta} - \Big[ 1+ a_{ij} \overline{\eta} +(M_f + H_0 L_f) \overline{\eta} \sum_{C_{hl}\in N_{r}(i,j)} C_{ij}^{hl} \Big] \\
&& \times \sup_{s \in (s_{k_q-1}, s_{k_q}]}  \big\| \phi_{\zeta} (s) - \phi_{\widetilde{\zeta}}(s) \big\|.
\end{eqnarray*}
The last inequality yields 
$
\displaystyle \sup_{s \in (s_{k_q-1}, s_{k_q}]}  \big\| \phi_{\zeta} (s) - \phi_{\widetilde{\zeta}}(s) \big\| > N_0,
$
where
$$
N_0 = \displaystyle \frac{\overline{\epsilon}_0 \underline{\eta}}{2+\displaystyle\overline{\eta} \max_{(i,j)}\Big[ a_{ij}  +(M_f + H_0 L_f)  \sum_{C_{hl}\in N_{r}(i,j)} C_{ij}^{hl} \Big]}.
$$

Now, let $\epsilon_1$ be the minimum of the numbers $\displaystyle \frac{\underline{\eta}}{2}$ and $\displaystyle\frac{N_0}{4\Big[M_F + H_0 \displaystyle \max_{(i,j)} \Big( a_{ij} + M_f  \sum_{C_{hl}\in N_{r}(i,j)} C_{ij}^{hl}  \Big) \Big]}.$ 

First of all, suppose that
$
\displaystyle \sup_{s \in (s_{k_q-1}, s_{k_q}]}  \big\| \phi_{\zeta} (s) - \phi_{\widetilde{\zeta}}(s) \big\| = \big\| \phi_{\zeta} (\nu) - \phi_{\widetilde{\zeta}}(\nu) \big\|
$
for some $\nu \in (s_{k_q-1},s_{k_q}],$ and denote 
$$
\gamma_q=\left\{\begin{array}{ll} \nu, & \textrm{if}~ \nu \le (s_{k_q-1}+s_{k_q})/2, \\
                                  \nu-\epsilon_1, &\textrm{if}~ \nu >(s_{k_q-1}+s_{k_q})/2, 
\end{array} \right..
$$
If $s$ belongs to the interval $\widetilde{J}_q=[\gamma_q, \gamma_q+\epsilon_1],$ then we have that 
\begin{eqnarray*}
&& \big|\phi^{ij}_{\zeta}(s)-\phi^{ij}_{\widetilde{\zeta}}(s)\big| \ge \big|\phi^{ij}_{\zeta}(\nu)-\phi^{ij}_{\widetilde{\zeta}}(\nu)\big| - \displaystyle \Big| \int_{\nu}^s a_{ij} \big|\phi^{ij}_{\zeta}(\tau)-\phi^{ij}_{\widetilde{\zeta}}(\tau)\big| d\tau \Big| \\
&& - \displaystyle \Big| \int_{\nu}^s   \sum_{C_{hl}\in N_{r}(i,j)} C_{ij}^{hl} \big| f\big( \phi_{\zeta}^{hl} (\tau) \big) \phi_{\zeta}^{ij}(\tau) - f\big( \phi_{\widetilde{\zeta}}^{hl} (\tau) \big) \phi_{\widetilde{\zeta}}^{ij}(\tau)  \big|  d\tau \Big|  \\
&& - \displaystyle \Big| \int_{\nu}^s    \big|   \zeta_{k_q}^{ij} - \widetilde{\zeta}_{k_q}^{ij} \big|  d\tau \Big| \\
&& \ge \big|\phi^{ij}_{\zeta}(\nu)-\phi^{ij}_{\widetilde{\zeta}}(\nu)\big| -2 \epsilon_1  \Big[ M_F+ H_0 \Big( a_{ij}  +M_f \sum_{C_{hl}\in N_{r}(i,j)} C_{ij}^{hl}\Big) \Big].
\end{eqnarray*} 
Hence, one can obtain that
\begin{eqnarray*}
\big\| \phi_{\zeta}(s)-\phi_{\widetilde{\zeta}}(s) \big\| > N_0 -2 \epsilon_1  \displaystyle \Big[ M_F+ H_0 \max_{(i,j)}\Big( a_{ij}  +M_f \sum_{C_{hl}\in N_{r}(i,j)} C_{ij}^{hl}\Big) \Big] \ge \displaystyle \frac{N_0}{2}.
\end{eqnarray*}
On the other hand, if  
$
\displaystyle \sup_{s \in (s_{k_q-1}, s_{k_q}]}  \big\| \phi_{\zeta} (s) - \phi_{\widetilde{\zeta}}(s) \big\| = \big\| \phi_{\zeta} (s_{k_q-1}+) - \phi_{\widetilde{\zeta}}(s_{k_q-1}+) \big\|,
$ then it can be shown in a similar way that the inequality $\big\| \phi_{\zeta}(s)-\phi_{\widetilde{\zeta}}(s) \big\| > \displaystyle \frac{N_0}{2}$ is valid also for $s \in \widetilde{J}_q=(s_{k_q-1},s_{k_q}].$ 

Therefore, $\big\| \varphi_{\zeta} (t) - \varphi_{\widetilde{\zeta}}(t) \big\| > \epsilon_0$ for $t \in \psi^{-1}(\widetilde{J}_q).$ Clearly, the intervals $J_q=\psi^{-1}(\widetilde{J}_q),$ $q\in \mathbb N,$ are disjoint. Consequently, the pair of bounded solutions $\varphi_{\zeta}(t),$ $\varphi_{\widetilde{\zeta}}(t) \in \mathscr{A}$ is frequently $(\epsilon_0,\epsilon_1)$-separated. $\square$

%%%%%%%%%%%%%% end of proof %%%%%%%%%%%%%%%

The main result of the present paper is as follows.

\begin{theorem}\label{main_SICNN_timescale}
If the conditions $(C1)-(C6)$ are valid and the map \eqref{timescale_mapF} is Li-Yorke chaotic on $\Lambda,$ then SICNN \eqref{SICNN_timescale} is Li-Yorke chaotic.
\end{theorem}

\noindent  \textbf{Proof.}
Let $\zeta=\left\{\zeta_k\right\}$ be a $q_0-$periodic solution of the map \eqref{timescale_mapF} for some natural number $q_0.$ In this case, the external input $L(t,\zeta)=\left\{L_{ij}(t,\zeta)\right\}$  is $p_0\omega-$periodic, where $p_0= \textrm{lcm} \left\{q_0,p\right\}/p.$ It can be shown that the bounded solution $\varphi_{\zeta}(t)$ of \eqref{SICNN_timescale} is $p_0\omega-$periodic. Therefore, there exists a $p_0\omega-$periodic solution of \eqref{SICNN_timescale} for each natural number $p_0.$

Denote by $\Sigma$ the set consisting of bounded solutions $\varphi_{\zeta}(t)$ of \eqref{SICNN_timescale} for which $\zeta_0 \in {\mathcal S},$ where ${\mathcal S} \subset \Lambda$ is the scrambled set of  \eqref{timescale_mapF}. One can confirm that the set $\Sigma$ is uncountable and no periodic solutions take place inside $\Sigma.$    

Using the Lemmas \ref{timescale_SICNN_prox} and \ref{SICNN_freq_separation_lemma}, one can confirm that any pair of different solutions inside $\Sigma$ is a Li-Yorke pair, i.e. $\Sigma$ is a scrambled set.  Lemma \ref{SICNN_freq_separation_lemma} implies also that for any solution $\varphi_{\zeta}(t) \in \Sigma$ and any periodic solution $\varphi_{\widetilde{\zeta}}(t) \in \mathscr{A},$ the pair $\varphi_{\zeta}(t),$ $\varphi_{\widetilde{\zeta}}(t)$ is frequently $(\epsilon_{0},\epsilon_{1})$-separated for some positive numbers $\epsilon_{0}$ and $\epsilon_{1}.$ Consequently, the network \eqref{SICNN_timescale} is Li-Yorke chaotic. $\square$

In the following section, we will present an example, which supports the result of Theorem \ref{main_SICNN_timescale}.

\section{An example} \label{SICNN_timescale_example_sec}
 
Consider the following SICNN,
\begin{eqnarray} \label{SICNN_exampe_timescale}
x^{\Delta}_{ij}(t)= -a_{ij} x_{ij}(t) -\displaystyle \sum_{C_{hl}\in N_{1}(i,j)} C_{ij}^{hl} f(x_{hl}(t)) x_{ij}(t) + L_{ij}(t,\zeta), ~ t\in \mathbb T_0,
\end{eqnarray} 
where $i,j=1,2,3,$ $\mathbb T_0 = \bigcup_{k=-\infty}^{\infty} [\theta_{2k-1}, \theta_{2k}],$ $\theta_k = \displaystyle \frac{7k}{4}  + \frac{1}{8} (9+(-1)^{k+1}),$ $k \in \mathbb Z,$ $f(s)=\displaystyle \frac{s^2}{5}$ for $\left|s\right| \le 1,$ $f(s)=\displaystyle \frac{1}{5}$ for $\left|s\right|>1$ and
$$\left( \begin{array}{ccc}
a_{11}&a_{12}&a_{13} \\
a_{21}&a_{22}&a_{23} \\
a_{31}&a_{32}&a_{33} \end{array} \right)= \left( \begin{array}{ccc}
1.9&1.5&2.1 \\
2.2&1.4&1.7 \\
1.8&1.6&2.3 \end{array} \right).$$ 
Notice that the time scale $\mathbb T_0$ satisfies the $\omega-$property with $\omega=7/2.$ In \eqref{SICNN_exampe_timescale}, for each $i$ and $j$ we will use the following coupling strengths,
$$  \left( \begin{array}{ccc}
C^{11}_{ij}&C^{12}_{ij}&C^{13}_{ij} \\
C^{21}_{ij}&C^{22}_{ij}&C^{23}_{ij} \\
C^{31}_{ij}&C^{32}_{ij}&C^{33}_{ij} \end{array} \right)= \left( \begin{array}{ccc}
0.003&0.001&0.005 \\
0.004&0.002&0 \\
0.006&0.001&0.003 \end{array} \right),$$ 
whenever the cells $C_{hl},$ $h,l=1,2,3,$ belong to $N_1(i,j).$ More precisely, for fixed $h$ and $l,$ the coupling strengths $C_{ij}^{hl},$ $i,j = 1,2,3,$ are taken to be equal to each other in cases where $C_{hl}$ belongs to $N_1(i,j).$ For instance, $C^{11}_{11}=C^{11}_{12}=C^{11}_{21}=C^{11}_{22}=0.003.$ Furthermore, for each $i$ and $j,$ we take $L_{ij}(t,\zeta)=\zeta_k$ for $t\in [\theta_{2k-1}, \theta_{2k}],$ $k\in\mathbb Z,$ where the sequence $\zeta=\left\{\zeta_k\right\}_{k\in\mathbb Z}$ is generated by the logistic map 
\begin{eqnarray}\label{timescale_example_map}
\zeta_{k+1}=\mu \zeta_k (1-\zeta_k).
\end{eqnarray}
The map \eqref{timescale_example_map} is chaotic in the sense of Li-Yorke for the values of the parameter $\mu$ between $3.84$ and $4$ \cite{Li75}. Moreover, for these values of the parameter, the interval $[0,1]$ is invariant under the iterations of the map \cite{Hale91}.   
 
\begin{figure*}[!ht] 
\centering
\includegraphics[width=15.0cm]{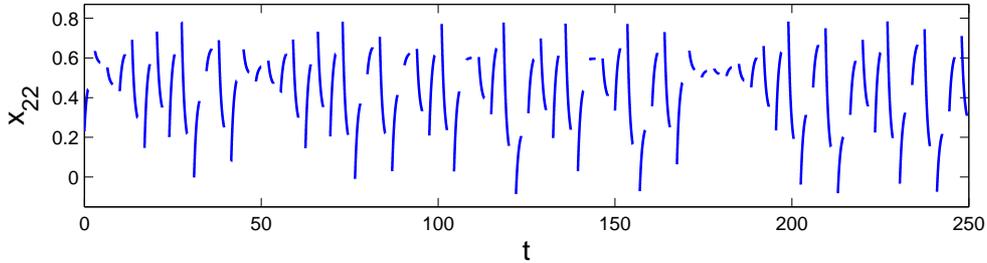}
\caption{Time series of the $x_{22}-$coordinate of SICNN \eqref{SICNN_exampe_timescale}. The figure manifests that the network behaves chaotically.}
\label{fig1}
\end{figure*}

Let us take $\mu=3.9$ in \eqref{timescale_example_map}. One can verify that the conditions $(C1)-(C6)$ are valid for  \eqref{SICNN_exampe_timescale} with $\delta=2,$ $p=1,$ $\psi(\omega)=3/2,$ $M_f=1/5,$ $L_f=2/5,$ $M_F=0.975,$ $\lambda_{11} \approx 1.2136,$ $\lambda_{12} \approx 1.0379,$ $\lambda_{13} \approx 1.3246,$ $\lambda_{21} \approx 1.3841,$ $\lambda_{22} \approx 1.0081,$ $\lambda_{23} \approx 1.1164,$ $\lambda_{31} \approx 1.1630$ $\lambda_{32} \approx 1.0744,$ $\lambda_{33} \approx 1.4460,$ $K_{11}=2.8,$ $K_{12}=2,$ $K_{13}=3.2,$ $K_{21}=3.4,$ $K_{22}=1.8,$ $K_{23}=2.4,$ $K_{31}=2.6,$ $K_{32}=2.2,$ $K_{33}=3.6,$ $H_0 \approx 10.7262,$ $\overline{c} \approx 0.1739, $ $\overline{d} \approx 0.2596.$ Thus, according to Theorem \ref{main_SICNN_timescale}, SICNN \eqref{SICNN_exampe_timescale} is Li-Yorke chaotic. We take $\zeta_0=0.715,$ and represent in Figure \ref{fig1} the $x_{22}$-coordinate of \eqref{SICNN_exampe_timescale} corresponding to the initial data $x_{11}(0)=0.42,$ $x_{12}(0)=0.35,$ $x_{13}(0)=0.56,$ $x_{21}(0)=0.48,$ $x_{22}(0)=0.23,$ $x_{23}(0)=0.39,$ $x_{31}(0)=0.48,$ $x_{32}(0)=0.25,$ $x_{33}(0)=0.51.$ Figure \ref{fig1} supports the result of Theorem \ref{main_SICNN_timescale} such that the network \eqref{SICNN_exampe_timescale} possesses Li-Yorke chaos.  
  
In the next section, we will demonstrate how to control the chaos of SICNN \eqref{SICNN_exampe_timescale} numerically by using the Pyragas control technique \cite{Pyragas92}.  
  
\section{Control of chaos} \label{control_SICNN_timescale} 

In the literature, control of chaos is understood as the stabilization of unstable periodic orbits embedded in a chaotic attractor. The studies on the control of chaos originated with Ott, Grebogi and Yorke \cite{Ott90}. The Ott-Grebogi-Yorke (OGY) control method depends on the usage of small time-dependent perturbations in an accessible system parameter to stabilize an already existing periodic orbit, which is initially unstable. Another well known technique for the control of chaotic systems is the Pyragas method \cite{Pyragas92}, which is known as delayed feedback control.

The source of the chaotic outputs generated by SICNN \eqref{SICNN_exampe_timescale} is the logistic map \eqref{timescale_example_map}. In this section, we will show that the chaos of \eqref{SICNN_exampe_timescale} can be controlled by applying the Pyragas control technique to the map \eqref{timescale_example_map}. More precisely, we will stabilize the unstable $7/2-$periodic solution of \eqref{SICNN_exampe_timescale} by performing the Pyragas method to \eqref{timescale_example_map} around the fixed point $1-1/ \mu$ of the map. 

The chaos control techniques such as the Pyragas and OGY methods are suitable for either continuous-time or discrete-time systems \cite{Pyragas92,Gon04,Ott90,Sch99}. However, to the best of our knowledge, there is no chaos control technique for dynamics on time scales. This is the main reason why we apply the Pyragas control to the logistic map in order to control the chaos of \eqref{SICNN_exampe_timescale}. It is worth noting that the OGY algorithm can also be applied to the logistic map to control the chaos of \eqref{SICNN_exampe_timescale}.

Now, let us describe the Pyragas control technique for the logistic map \eqref{timescale_example_map}. The aim in this method is to stabilize the unstable fixed point $1-1/ \mu$ of \eqref{timescale_example_map} by using the delayed feedback control algorithm
\begin{eqnarray} \label{SICNN_control_eq0}
\overline{\zeta}_{k+1} = \mu \overline{\zeta}_k (1-\overline{\zeta}_k) + \beta (\overline{\zeta}_k - \overline{\zeta}_{k-1}),
\end{eqnarray}
where $\beta$ is the feedback gain \cite{Pyragas92,Pyragas10,Zelinka}. According to the results of the paper \cite{Pyragas10}, the nonzero fixed point becomes stable for 
$
\displaystyle \frac{\mu-3}{2}<\beta<1,
$
and the optimal value of the feedback gain, which leads to the fastest convergence of nearby initial conditions is $\beta_{op}=\mu - 2 (\mu-1)^{1/2}.$

It was mentioned in \cite{Pyragas10}  that the algorithm \eqref{SICNN_control_eq0} is successful if the initial data are sufficiently close to the fixed point, and solutions of \eqref{SICNN_control_eq0} can escape to infinity for some initial data from the chaotic attractor of \eqref{timescale_example_map}. In order to guarantee the stabilization for any initial data, Pyragas and Pyragas  \cite{Pyragas10}  proposed that the condition
\begin{eqnarray} \label{SICNN_control_eq1}
|\overline{\zeta}_{k}-\overline{\zeta}_{k-1} | < \varepsilon
\end{eqnarray}
has to be satisfied for a suitably chosen small positive number $\varepsilon,$ which can be determined only numerically. The inequality \eqref{SICNN_control_eq1} is used to estimate the strength of the perturbation $\beta (\overline{\zeta}_k - \overline{\zeta}_{k-1})$ if it would be applied at a given iteration step $k.$ On the other hand, the condition
\begin{eqnarray} \label{SICNN_control_eq2}
\overline{\zeta}_{k-1} > \displaystyle \frac{\mu-1- [(1-\mu)^2-4\mu \varepsilon]^{1/2}}{2\mu}
\end{eqnarray}
is required to avoid the stabilization of the zero fixed point \cite{Pyragas10}. 

In the Pyragas control method, at each iteration step $k$ after the control mechanism is switched on, we consider the map \eqref{SICNN_control_eq0} with a chosen value of $\beta$ between $(\mu-3)/2$ and $1$ provided that \eqref{SICNN_control_eq1} and \eqref{SICNN_control_eq2} are valid. Otherwise, we take $\beta=0$ and wait until both \eqref{SICNN_control_eq1} and \eqref{SICNN_control_eq2} are satisfied. If this is the case, the control of chaos is not achieved
immediately after switching on the control mechanism. Instead, there is a transition time that depends on the ergodic properties of \eqref{timescale_example_map} before the fixed point is stabilized.

Now, we will apply the Pyragas control technique to the logistic map \eqref{timescale_example_map} with $\mu = 3.9$ around the fixed point $2.9/3.9$ to control the chaos of \eqref{SICNN_exampe_timescale}.  For that purpose, we take into account the following auxiliary network, 
\begin{eqnarray} \label{SICNN_control_network}
z^{\Delta}_{ij}(t)= -a_{ij} z_{ij}(t) -\displaystyle \sum_{C_{hl}\in N_{1}(i,j)} C_{ij}^{hl} f(z_{hl}(t)) z_{ij}(t) + L_{ij}\big(t,\overline{\zeta}\big), ~ t\in \mathbb T_0,
\end{eqnarray} 
where $i,j=1,2,3$ and $\overline{\zeta}=\left\{\overline{\zeta}_k\right\}_{k \in \mathbb Z}$ is a solution of (\ref{SICNN_control_eq0}). The network \eqref{SICNN_control_network} is the control network conjugate to \eqref{SICNN_exampe_timescale}. The time scale $\mathbb T_0,$ activation function $f,$ passive decay rates $a_{ij}$ and coupling strengths $C_{ij}^{hl}$ in \eqref{SICNN_control_network} are the same with the ones used in \eqref{SICNN_exampe_timescale}. For each $i$ and $j,$ the external inputs $L_{ij}(t,\overline{\zeta})$ in \eqref{SICNN_control_network} are defined as $L_{ij}(t,\overline{\zeta})=\overline{\zeta}_k$ for $t\in [\theta_{2k-1}, \theta_{2k}],$ $k\in\mathbb Z.$

\begin{figure*}[!ht] 
\centering
\includegraphics[width=15.0cm]{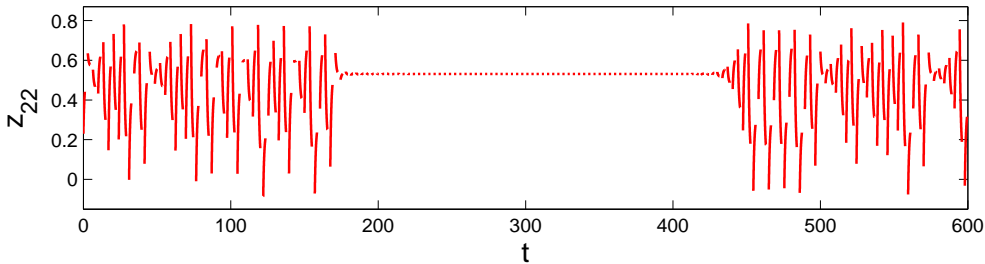}
\caption{Stabilization of the unstable $7/2-$periodic solution of SICNN \eqref{SICNN_exampe_timescale} by means of the conjugate control network \eqref{SICNN_control_network}. The chaos of \eqref{SICNN_exampe_timescale} is controlled by applying the Pyragas control technique to the logistic map \eqref{timescale_example_map} around the fixed point $2.9/3.9.$ The values $\beta=\beta_{op} \approx 0.4941$ and $\varepsilon = 0.025$ are utilized in the simulation. The control mechanism is switched on at $t=139.5$ and switched off at $t=349.5.$}
\label{fig2}
\end{figure*}

Figure \ref{fig2} shows the $z_{22}$-coordinate of \eqref{SICNN_control_network} corresponding to the initial data $z_{11}(0)=0.42,$ $z_{12}(0)=0.35,$ $z_{13}(0)=0.56,$ $z_{21}(0)=0.48,$ $z_{22}(0)=0.23,$ $z_{23}(0)=0.39,$ $z_{31}(0)=0.48,$ $z_{32}(0)=0.25,$ $z_{33}(0)=0.51$ and $\overline{\zeta}_0=0.715.$ The control mechanism is switched on at $t=\theta_{79}=139.5$ and switched off at $t=\theta_{199}=349.5.$ In the simulation, we take $\varepsilon = 0.025,$ and make use of the optimum value $\beta_{op} \approx 0.4941$ of the feedback gain $\beta.$ It is seen in Figure \ref{fig2} that the unstable $7/2-$periodic solution of \eqref{SICNN_exampe_timescale} is stabilized. The control becomes dominant approximately at $t=216.5$ and its effect prolongs approximately till $t=414,$ after which chaos emerges again and irregular behavior reappears. It is worth noting that the stabilization does not occur immediately after switching on the control mechanism, but rather there is a transition time. The simulation result demonstrates that it is sufficient to stabilize an unstable periodic orbit of the logistic map \eqref{timescale_example_map} to control the chaos of SICNN \eqref{SICNN_exampe_timescale}. 

\section{Conclusions} \label{conc_section_SICNN_timescales}

The present study is concerned with the existence of Li-Yorke chaos in SICNNs on time scales. We rigorously prove the ingredients of Li-Yorke chaos, proximality and frequent separation, by taking advantage of the external inputs. This is the first time in the literature that chaos is obtained for neural networks on time scales. Our results reveal that external inputs are capable of generating chaotic outputs on time scales. The presented technique can be used to obtain chaotic SICNNs on time scales without any restriction in the number of cells. 

Another significant result of our paper is the control of the chaos on time scales. The presence of unstable periodic motions embedded in the chaotic attractor is evidenced by means of the Pyragas control technique \cite{Pyragas92,Pyragas10}. It is numerically demonstrated that the Pyragas method is appropriate for controlling chaos not only in discrete and continuous-time systems, but also in systems on time scales. 

The detection of chaotic behavior in neural networks is of great interest since it may provide the opportunity to understand how the brain and the rest of the nervous system works. We believe that the presented technique will shed light on the mathematical investigation of neural processes that work intermittently \cite{Dowling01,Perrins02,Evans02}. Moreover, our results can provide further research areas in neural activities that are achievable at certain time intervals. The problem of period-doubling route to chaos \cite{Feigenbaum80,Sander11} by neural networks on time scales can be considered in the future through the presented method. Our approach can also be useful for designing secure communication systems \cite{Cheng13,Khadra03,Muthukumar13,Muthukumar14}, and it can be improved for other kinds of recurrent networks such as Hopfield and Cohen-Grossberg neural networks \cite{Hopfield84,Cohen93}.

\section*{Acknowledgments}

This work is supported by the 2219 scholarship programme of T\"{U}B\.{I}TAK, the Scientific and Technological Research Council of Turkey.

%%%%%%%%%%%%%%%%%%%%% References %%%%%%%%%%%%%%%%%%%%%%%%%

%\section*{References}

%\bibliographystyle{model1-num-names}
%\bibliography{elsarticle-num.bst}

\end{document}